\documentclass[dvipsnames]{article}

\usepackage[margin=1in]{geometry}

\usepackage{graphicx,amsmath,upgreek,bm}
\usepackage{colortbl}
\usepackage{wrapfig}
\usepackage[rgb, table]{xcolor}
\usepackage{booktabs}
\usepackage{makecell}
\usepackage{pifont}
\usepackage{todonotes}
\usepackage[toc,page]{appendix}
\usepackage{multirow}
\usepackage{algorithm}
\usepackage{rotating}
\usepackage{hyperref}
\hypersetup{
   colorlinks=true,
   linkcolor=blue,
   filecolor=magenta,
   urlcolor=blue,
   citecolor=blue,
	}
\usepackage[
  separate-uncertainty = true,
  multi-part-units = repeat
]{siunitx}
\usepackage{caption} 
\usepackage{subcaption}
\usepackage{amsmath}
\usepackage{gensymb}
\usepackage{mathtools}
\usepackage{amssymb}  
\usepackage{appendix}    
\usepackage{booktabs} 
\usepackage[para,online,flushleft]{threeparttable}
\usepackage{textcomp}
\usepackage[noend]{algpseudocode}
\usepackage{stackengine}
\usepackage{arydshln}
\makeatletter
  \renewcommand*\env@matrix[1][*\c@MaxMatrixCols c]{%
    \hskip -\arraycolsep
    \let\@ifnextchar\new@ifnextchar
  \array{#1}}
\makeatother
\makeatletter
\def\BState{\State\hskip-\ALG@thistlm}
\makeatother
\definecolor{light-gray}{gray}{0.9}

\title{Chatter Detection in Turning Using Machine Learning and Similarity Measures of Time Series via Dynamic Time Warping}

\author{Melih C. Yesilli\\
				Department of Mechanical Engineering\\
				Michigan State University\\
				yesillim@egr.msu.edu
			\and
				Firas A.~Khasawneh\\
				Department of Mechanical Engineering\\
				Michigan State University\\
				khasawn3@egr.msu.edu
			\and
				Andreas Otto\\
				Institute of Physics\\
				Chemnitz University of Technology\\
				andreas.otto@physik.tu-chemnitz.de
				}
				
\date{}

\begin{document}
\maketitle

\begin{abstract}
Chatter detection from sensor signals has been an active field of research.
While some success has been reported using several featurization tools and machine learning algorithms, existing methods have several drawbacks such as manual preprocessing and requiring a large data set.
In this paper, we present an alternative approach for chatter detection based on K-Nearest Neighbor (kNN) algorithm for classification and the Dynamic Time Warping (DTW) as a time series similarity measure.
The used time series are the acceleration signals acquired from the tool holder in a series of turning experiments. 
Our results, show that this approach achieves detection accuracies that in most cases outperform existing methods. We compare our results to the traditional methods based on Wavelet Packet Transform (WPT) and the Ensemble Empirical Mode Decomposition (EEMD), as well as to the more recent Topological Data Analysis (TDA) based approach. We show that in three out of four cutting configurations our DTW-based approach attains the highest average classification rate reaching in one case as high as  $99\%$ accuracy. Our approach does not require feature extraction, is capable of reusing a classifier across different cutting configurations, and it uses reasonably sized training sets. Although the resulting high accuracy in our approach is associated with high computational cost, this is specific to the DTW implementation that we used. Specifically, we highlight available, very fast DTW implementations that can even be implemented on small consumer electronics.  Therefore, further code optimization and the significantly reduced computational effort during the implementation phase make our approach a viable option for in-process chatter detection.

% The computation cost is also highest during the training phase, so further code optimization and the significantly reduced computational effort during the implementation phase make our approach a viable option for in-process chatter detection. 
% Our approach relies on using k-NN in combination with distance matrices that are directly constructed from the signals using DTW and DFD; thus, they do not require any user-dependent pre-processing for feature extraction. 
% Further, we show that the DTW approach allows training on one cutting configuration and testing on another as evidenced by the transfer learning results which significantly exceed their WPT and EEMD counterparts. 
% The similarity-based approach also allows distinguishing between chatter and intermediate chatter which look different in the time domain, but may look similar in the frequency domain. 
% Distinguishing between these two types of chatter is not possible with the WPT EEMD methods when the signals have similar frequency contents. 
% In contrast to black-box machine learning tools, all of our results are obtained using reasonably sized training and testing sets. 
\end{abstract}


\section{Introduction}
\label{sec:intro}
Machine tool chatter manifests as excessive vibrations of the cutting tool or the workpiece during machining.  
This phenomenon often leads to poor surface finish, and it can shorten the lifetime of the cutting tools. 
Therefore, various methods for chatter prediction and mitigation techniques have been proposed in the past several decades. 
Munoa et al. describes several different chatter mitigation techniques including increasing stiffness of the machine tools, passive and active damping techniques and changing the spindle speed \cite{Munoa2016}.
While chatter prediction tools are important for process planning, the cutting parameters often drift during the process which necessitates utilizing sensor signals to detect chatter in a practical setting.  
Traditionally, chatter detection tools combine methods from signal processing, such as Wavelet Packet Transform (WPT) and Empirical Mode Decomposition (EMD) or Ensemble Empirical Mode Decomposition (EEMD), with machine learning algorithms. 
While SVM is the most widely adopted classification algorithm for chatter detection in the literature, other classifiers for chatter diagnosis that have been successfully used include Neural network classification \cite{Lamraoui2013}, logistic regression \cite{Ding2017}, Quadratic Discriminant Analysis \cite{Thaler2014} and Hidden Markov Model \cite{Han2016}. 

For example, Chen and Zheng studied online chatter detection in milling using WPT and SVM with RFE \cite{Chen2017}. 
Ji et al.~extracted informative IMFs from accelerometer signals in milling \cite{Ji2018}. 
These IMFs were used to reconstruct a signal from which the standard deviation, power spectral entropy, and the fractal dimension were used as features in an SVM chatter classifier. 
Liu et.~al combined Wavelet Packet Decomposition (WPD) and EMD to obtain informative IMFs with reduced modal aliasing effects from cutting force signals in milling \cite{Liu2017}. 
These IMFs were used to reconstruct the signal and obtain its Hilbert-Huang Transform (HHT). 
The mean value and the standard deviation of the spectrum of the resulting HHT were used to set a threshold for chatter. 
Chen et al.~used Fisher Discriminant Ratio for ranking of features obtained with EEMD \cite{Chen2018}. 
However, their approach requires tedious pre-processing, and the resulting thresholds are relatively too close. 

Although prior studies have shown some success with chatter detection using WPT and EEMD, these tools share two main limitations: 
(1) training a classifier requires significant pre-processing of the data, 
and (2) the trained classifier is sensitive to the differences between the training set and the test set \cite{Yesilli2019}. 
The pre-processing of the data first requires decomposing the signal into wavelet packets (for WPT) and IMFs (for EEMD).  
Manual analysis of the resulting decompositions is then necessary in order to identify the informative wavelet packets or informative IMFs. 
These are the parts of the decomposition that can inform the classifier about the existence of chatter in the signal. 
Both informative packets and informative IMFs are determined by examining the signal's Fourier spectrum by a skilled user and choosing the part of the signal decomposition that falls within the range of the chatter frequency. 
These informative functions are then used to reconstruct the time domain signal which allows extracting frequency and time features for chatter classification. 
The resulting features are often ranked with Recursive Feature Elimination (RFE) method, and they are combined with Support Vector Machine algorithm (SVM) to obtain a classifier. 
Once a classifier is obtained, an incoming data stream can be classified. 

However, it was shown in \cite{Yesilli2019} that both WPT and EEMD classifiers suffer from a significant decrease in accuracy if the cutting conditions shift.  
Specifically, Yesilli et al.~assessed the transfer learning performance of each of WPT and EEMD \cite{Yesilli2019}. 
In that work, given two configurations with two different chatter frequencies, an SVM classifier was trained with one configuration and tested on the other. 
It was shown that EEMD retains higher accuracy rates than WPT, thus making it a better approach in applications where the machine stiffness significantly varies during operation. 
These limitations require that WPT and EEMD be applied to cutting settings where the natural frequencies of the machine-tool system do not shift too much during the cutting process. 
Further, the level of skill required precludes these tools from being widely applied. 

In addition to WPT and EEMD, frequency spectrum analysis is also used for feature matrix generation. 
Thaler et al.~proposed chatter diagnosis methods based on the analysis of sound signals in band sawing processes using Short-Time Fourier Transform \cite{Thaler2014}. 
Lamraoui et al.~applied multi-band resonance filtering and envelope analysis to milling vibration signals \cite{Lamraoui2013}. 
Wang et al.~used the power spectrum and the $Q$-factor as a descriptor of chatter, and they combined these features with SVM \cite{Wang2018}. 
Variational Mode Decomposition (VMD) is also another method for chatter detection. 
For example, Liu et al.~developed a method to automatically select the VMD parameters, and to extract the corresponding features using signal energy entropy. 
Similar to WPT and EEMD, the above methods require expert users to build the chatter classification algorithm, and the resulting algorithm is highly specialized to the process generating the data. 
Cherukuri et al.~used Artificial Neural Network (ANN) on synthetic turning data for chatter classification \cite{Cherukuri2019}. 
However, using ANN (or other black-box machine learning methods) requires large training sets, that may not be always available especially in small-batch production, which constitutes a large portion of discrete manufacturing. 

Other, more recent methods to extract features from cutting vibration signals are based on Topological Data Analysis (TDA) \cite{Khasawneh2014,Khasawneh2016,Khasawneh2018}. 
For example, Yesilli et al.~\cite{Yesilli2019a} studied the chatter classification accuracy in turning using features extracted from persistence diagrams including persistence landscapes, persistence images, Carlsson Coordinates \cite{Adcock2016}, template functions \cite{Perea2019}, a kernel method \cite{Reininghaus2014}, and path signatures of persistence landscapes\cite{Chevyrev2018}. 
%\todo[inline]{To Melih: please add references to each of these TDA methods.} 
Using an SVM algorithm to train a classifier, Yesilli et al.~found that topological features do encode chatter signatures with Carlsson coordinates and template functions yielding the highest overall accuracy rates. 
The generation of the persistence diagrams can be largely automated, and the features are pulled directly from the resulting diagrams without requiring any user expertise. 
However, the transfer learning capabilities of these tools have not been tested, and reducing their computational time is still a topic of current research \cite{Bauer2017}. 
Therefore, there is a need for an accurate machine learning algorithm for chatter diagnosis that can 
(1) be easily and widely applied, and
(2) has good transfer learning capabilities,
(3) can be computed in reasonable time. 

In this paper, we describe a novel method for chatter identification that satisfies these conditions based on combining the K-Nearest Neighbor (kNN) classifier with time series similarity measure: Dynamic Time Warping (DTW).
DTW has been used in many application domains including speech recognition \cite{Myers1980,Myers1981,sakoe1990,Juang1984,Itakura1975}, time series classification \cite{Niennattrakul2007,Yu2007,Petitjean2014}, and signature verification \cite{Shanker2007,Munich1999,Parizeau1990,Martens1996}. 
In contrast to WPT and EEMD, no pre-processing is necessary since DTW operates on the original signals, and it does not require the signals to have the same lengths. 
We test the approach using a set of turning experiments where the cutting tool is instrumented with accelerometers. 
A total of four cutting configurations are used, and for each configuration a variety of depths of cut and cutting speeds are tested. 
We establish a criteria for tagging the resulting signals as chatter-free, intermediate chatter, or full chatter, see Section \ref{sec:data_tagging} for more details.
We then split the data in each configuration into training and testing sets, compute the pairwise distance matrices in the training set using DTW, then train an K-Nearest Neighbor classifier and use it to tag the test signals as chatter/chatter-free. 
We repeat the above process ten times where each time the data is randomly split into training/testing sets and we report the average test accuracy and standard deviation.  
We compare the resulting success rates to their counterparts in the literature including WPT, EEMD, as well as the most prominent TDA featurization tools. 
Our results show that in three out of the four cutting configurations DTW scores the highest average accuracy. 
Further, we show that, in contrast to WPT and EEMD, DTW have excellent transfer learning capabilities. 
Finally, we further show that DTW can successfully distinguish between three classes: chatter, intermediate chatter, and no-chatter. 
In general, using WPT and EEMD for distinguishing chatter from intermediate chatter is difficult because although the time-domain signals in these two cases are different, their corresponding frequency spectra are too close. 
Therefore, our approach enables further classification beyond just chatter and no chatter. 

The paper is organized as follows. 
Section \ref{sec:experimental_setup} describes the experimental setup. 
Section \ref{sec:dtw_and_dfd} details our approach which combines similarity measure of time series with machine learning for chatter detection. 
Section \ref{sec:k-Nearest-Neighbor} briefly explains the K-Nearest Neighbor algorithm.
Section \ref{sec:results} provides the results and compares the accuracies and runtime of our approach against the WPT, EEMD, and topological feature extraction methods, while our concluding remarks can be found in Sec.~\ref{sec:conclusion}.

\section{Experimental Setup and Data Preprocessing}
\label{sec:experimental_setup}
%************************************
This section describes the experimental setup, the preprocessing of the experimental data, and the data. 
\subsection{Experimental Setup}
\begin{figure}[h]
\centering
\includegraphics[width=0.70\textwidth,height=.75\textheight,keepaspectratio]{experimental_setup.png}
\hspace{0.25in}
\includegraphics[width=0.185\textwidth]{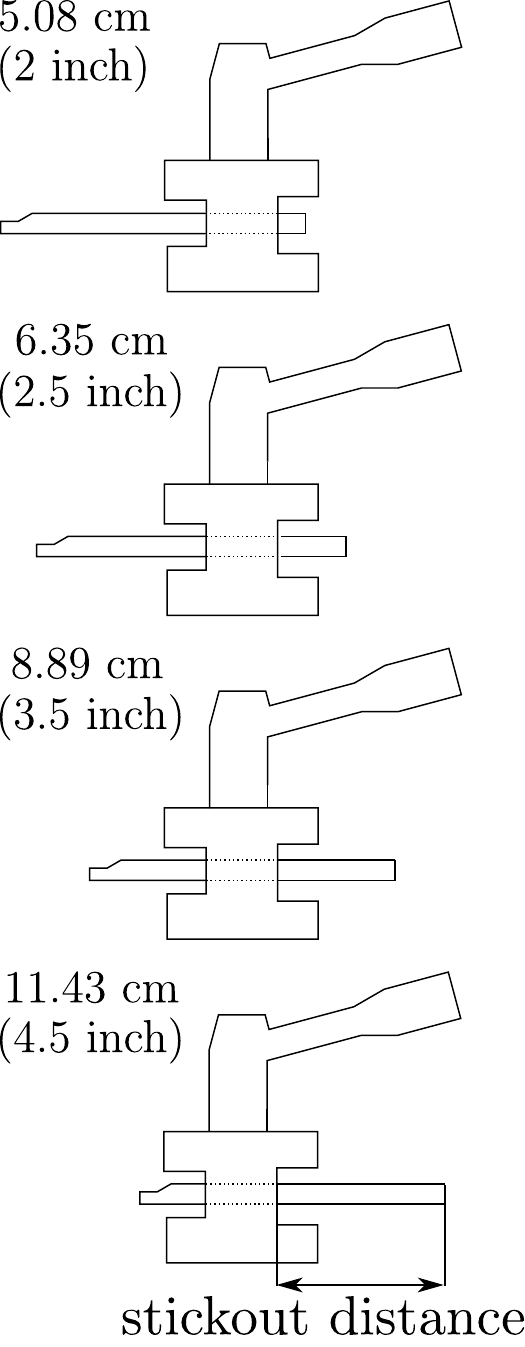}
\caption{The experimental setup (left), and an illustration of the four stickout lengths (right).}
\label{fig:experimental_setup}
\end{figure}

Figure \ref{fig:experimental_setup} shows the experimental setup for the turning cutting tests. 
A Clausing-Gamet $33$ cm ($13$ inch) engine lathe was utilized and it was equipped with three accelerometers to measure vibration signals: 
two PCB 352B10 uniaxial accelerometers and a PCB 356B11 triaxial accelerometer. 
The uniaxial accelerometers were attached to an S10R-SCLCR3S boring bar where the latter is part of Grizzly T10439 carbide insert boring bar set.  
Cutting was performed using $0.04$ cm (0.015 inch) radius Titanium nitride coated insert. 
The distance between the uniaxial accelerometers and the cutting insert was set to $3.81$ cm ($1.5$ inch) to protect the accelerometers from cutting debris. 
The vibration signals were collected at $160$ kHz using an NI USB-6366 data acquisition box and Matlab's data acquisition toolbox. 

%---------------------------------
%---------------------------------

Experiments were performed for four different cutting configurations based on the stickout distance which is the distance between back side of the cutting tool holder and the hell of the boring bar, see Fig.~\ref{fig:experimental_setup}. 
Changing the stickout distance varies the stiffness of the boring rod, and consequently the corresponding chatter frequencies. 
The four stickout lengths are: $5.08$ cm ($2$ inch), $6.35$ cm ($2.5$ inch), $8.89$ cm ($3.5$ inch) and $11.43$ cm ($4.5$ inch). 
For each stickout length, we performed cutting tests using several combinations of spindle speeds and depths of cut. 
Moreover, A PCB 130E20 microphone and Terahertz LT880 laser tachometer was also utilized to collect acoustics signals and rotational rate signals of spindle, respectively.

\subsection{Preprocessing of Experimental Data}
The sampling frequency for the experiment was 160 kHz. 
The reason for oversampling to is that we did not use an in-line analog filter during experiments. 
Instead, to avoid the aliasing effects, Butterworth low pass filter of order $100$ was used, and the data was subsequently downsampled to $10$ kHz. 
The filtered and downsampled data was used to label the data as explained in Sec.~\ref{sec:data_tagging}. 
Both the raw and filtered data are available for download in a Mendeley repository \cite{Khasawneh2019}.

\subsection{Data Tagging}
\label{sec:data_tagging}

\begin{figure}[H]
\centering
\includegraphics[width=0.9\textwidth,height=.9\textheight,keepaspectratio]{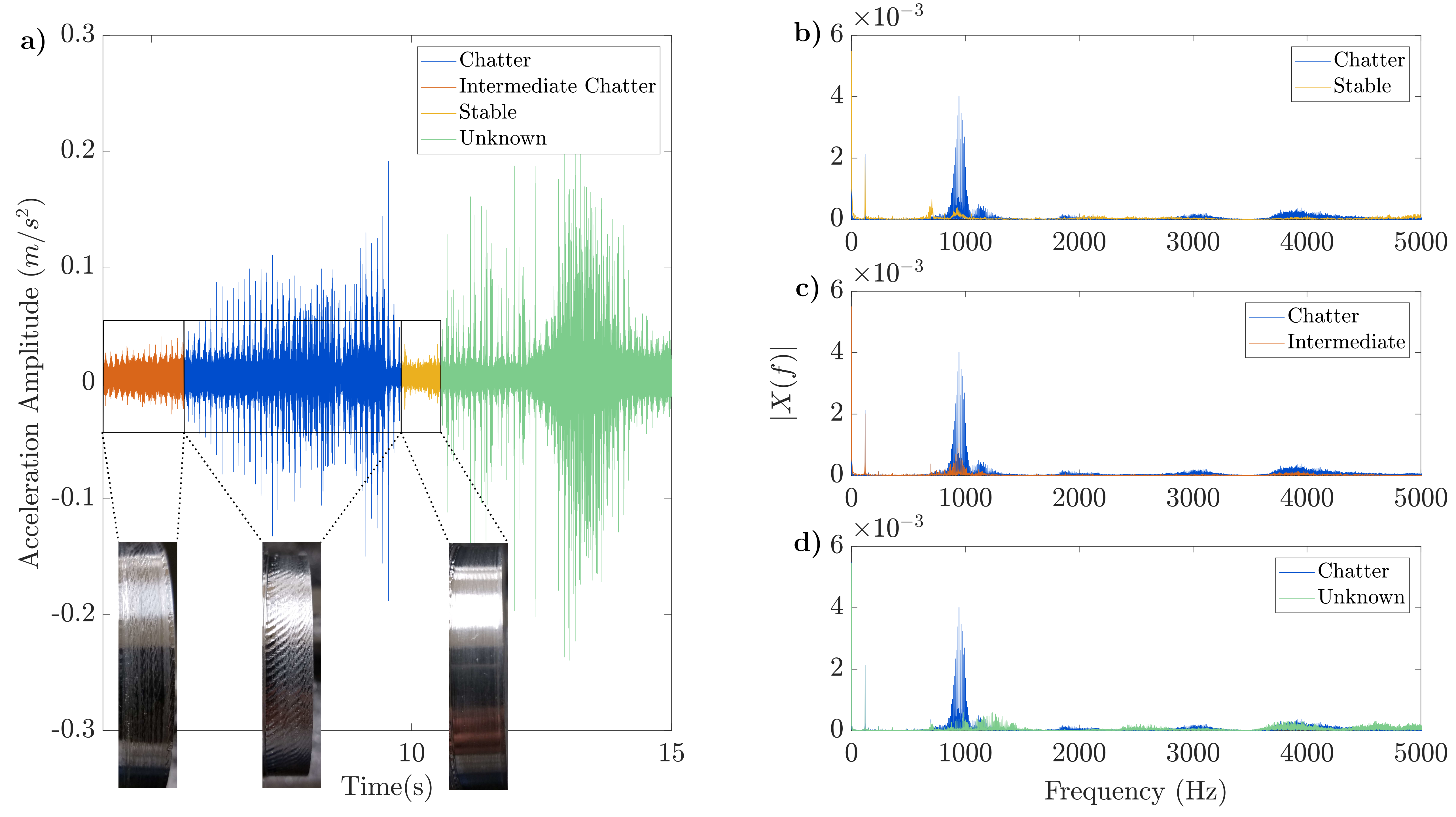}
\caption{(a) Sample tagged time series and some samples of the resulting surface finish. 
The right panel shows a comparison of the frequency spectra between a signal labeled as chatter versus (b) no chatter, (c) intermediate chatter, and (d) unknown.}
\label{fig:tagged_data}
\end{figure}

While several sensors were used to collect data during cutting, only the $x$-axis vibration signal of the tri-axial accelerometer was used. 
This is because most of the acceleration signals obtained from the other accelerometers are redundant, and the $x$-axis vibration signals of tri-axial accelerometer had the best Signal-to-Noise-Ratio (SNR). 
The data was labeled using four different tags:  
stable or chatter-free (s), intermediate chatter (i), chatter (c) and unknown (u). 
Note that the same cutting signal can contain regions with different labels as shown in Fig.~\ref{fig:tagged_data}. 

%---------------------------------
%---------------------------------

When tagging the time series, both time and frequency domain characteristics were considered. 
In the time domain, the raw data was divided according to its amplitude. 
In the frequency domain, we only considered frequencies below $5$ kHz. 
Using this information, tagging was performed by looking at the peaks in time and frequency domain. 
If the signal has low peaks in both domains, then it is labeled as stable or chatter-free (s). 
These time series also have high peaks at the spindle rotation frequency (see Fig.\ref{fig:tagged_data}b) \cite{Insperger2008}. 
Time series with intermediate chatter have low peaks in the time domain, but they have high peaks in the frequency domain. 
On the other hand, a time series was labeled as chatter signal if it had high peaks in both domains.  
Any signal that did not fit any of the above criteria was labeled as unknown (u). 
The resulting tags were verified by checking the corresponding surface finish of the workpiece as shown using the example photos in Fig.~\ref{fig:tagged_data}. 
Also, the number of time series belongs to each label for all stickout cases is provided in Tab.\ref{tab:time_series_number}.

%----------------
\begin{table}[ht]
% \rowcolors{8}{gray!30}{white}
\centering
	\begin{tabular}{c|c|c|c|c}
		\makecell{Stickout length\\ (cm (inch))} & \makecell{ Stable} & \makecell{Mild chatter} & \makecell{Chatter }  & \makecell{Total} \\
		\toprule 
		5.08 (2)	  & 17 	& 8   & 11 & 36 \\
		6.35 (2.5)	& 7 	& 4   & 3  & 14 \\
		8.89 (3.5)  & 7 	& 2   & 2  & 11 \\
		11.43 (4.5) & 13 	& 4   & 5  & 22 \\
	\end{tabular}
	\caption{Number of time series in each label for all stickout cases.}
	\label{tab:time_series_number}
\end{table}
%----------------

\section{Similarity-based method for chatter detection using Dynamic Time Warping (DTW)}
\label{sec:dtw_and_dfd}
%************************************

This section describes a novel method for chatter detection using the similarity between time series. The metric we use for measuring similarity is the Dynamic Time Warping (DTW). We first define DTW in Section \ref{sec:dtw_review}. We then describe how DTW generate similarity matrices which can be used for machine learning in Section \ref{sec:process}.

%************************************
\subsection{Dynamic Time Warping(DTW)}
\label{sec:dtw_review}
\begin{figure}[h]
\centering
\includegraphics[width=0.65\textwidth,height=.95\textheight,keepaspectratio]{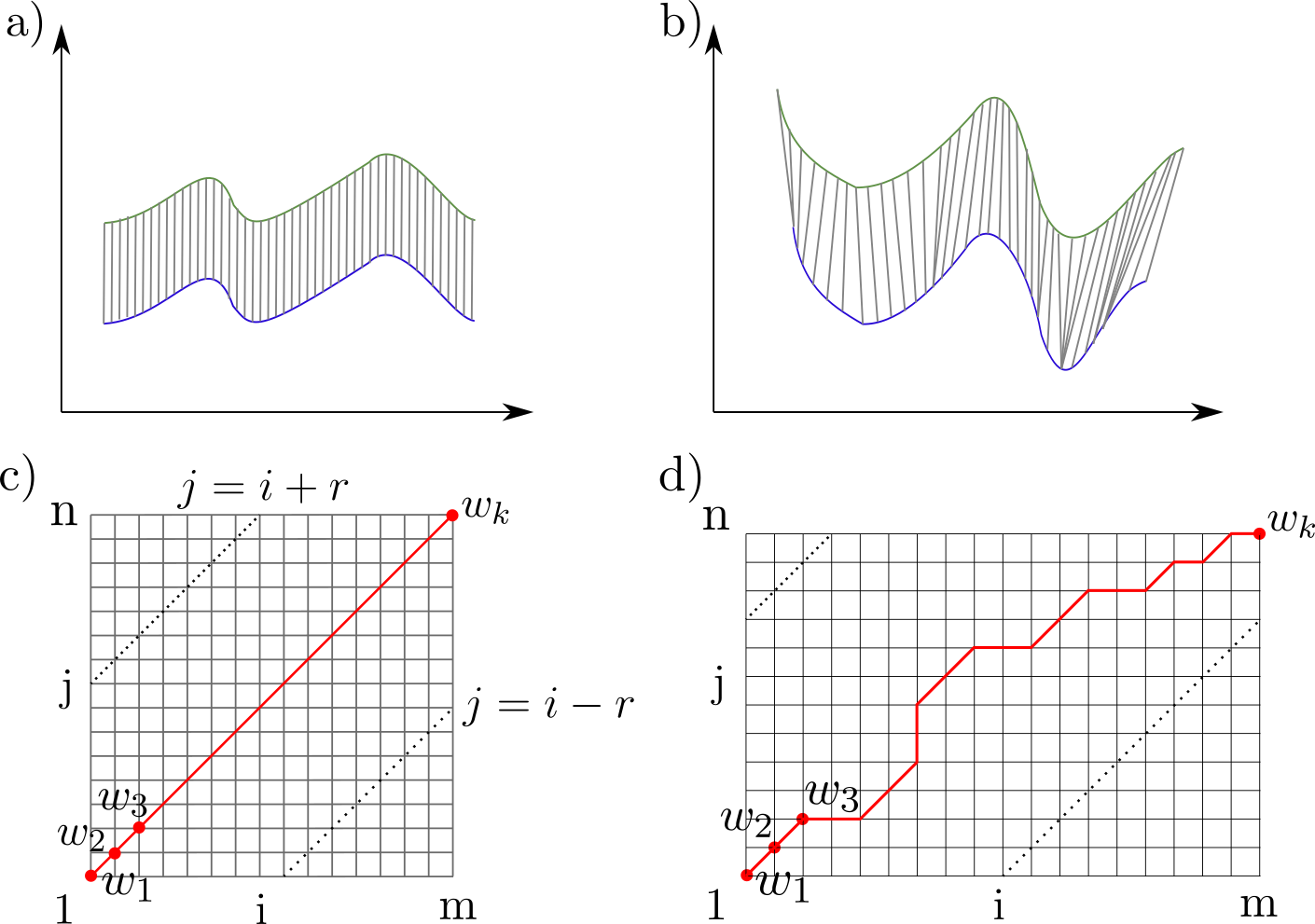}
\caption{DTW alignment and warping path for time series with identical length (a,c) and for time series with different length (b,d).}
\label{fig:warping_paths}
\end{figure}

%************************************
Dynamic Time Warping is an algorithm which is capable of measuring distance or similarity between two time series even if they have dissimilar lengths. Let $TS_{1}$ and $TS_{2}$ be two time series with elements $x_i$ and $y_j$ whose lengths are $m$ and $n$ as follows:
%---------------------------------
\begin{gather*}
TS_{1} = x_{1},x_{2},\ldots,x_{i},\ldots,x_{m},\\
TS_{2} = y_{2},y_{2},\ldots,y_{j},\ldots,y_{n}.
\end{gather*}
%---------------------------------
Berndt and Clifford \cite{berndt1994} state that the warping path $w_{k}=(x_{i(k)},y_{j(k)})$ between two time series can be represented by mapping the corresponding elements of the time series on $m \times n$ matrix (see Fig.~\ref{fig:warping_paths} for warping path examples). The warping path is composed of the points $w_{k}$ which indicate alignment between the elements $x_{i(k)}$ and $y_{j(k)}$ of the time series. The length $L$ of the warping path fulfills the constraints $m \leq L \leq n$, where we assume that $n \geq m$. For instance, $w_{3}$ in Fig.~\ref{fig:warping_paths}d corresponds to the alignment of $x_{3}$ and $y_{2}$. In the case $m=n$, the warping path is always the diagonal line through the $m\times n$ matrix, i.e. $w_{k}=(x_{k},y_{k})$ (see Fig.~\ref{fig:warping_paths}c). However, in general, warping paths are not unique and several warping paths can be generated for the same two time series. For dissimilar length the DTW algorithm chooses the warping path that gives the minimum distance between the element pairs under certain constraints. While there are several options for computing the distance between a pair $(x_i,y_j)$ of elements of the time series, in this implementation, the Euclidean distance $d(x_{i},y_{j})=\left\lVert x_i-y_j \right\lVert_2$ is used. The minimization of the distance between $TS_1$ and $TS_2$ in the DTW algorithm can then be written according to \cite{berndt1994}.
%---------------------------------
\begin{equation}
D_{\rm TW}(TS_{1},TS_{2}) = min\left(\sum\limits_{k=1}^L d(w_{k})\right).
\end{equation} 
%---------------------------------

%---------------------------------
%---------------------------------

There are several restrictions to define disparate warping paths. These are monotonicity, continuity, adjustment window condition, slope constraint and boundary conditions. These restrictions are applied on the alignment window to reduce the possible number of warping paths since there is excessive number of possibilities for warping paths without any constraint \cite{Sakoe1978}.
\begin{itemize}
	\item Monotonicity : The indices $i$ and $j$ should always either increase or stay the same such that $i(k)\geq i(k-1)$ and $j(k)\geq j(k-1)$.
	\item Continuity: The indices $i$ and $j$ can only increase at most by one such that $i(k)-i(k-1) \leq 1$ and $j(k)-j(k-1) \leq 1$.
	\item Boundary condition: The warping paths should start where $i$ and $j$ are equal to $1$ and should end where $i=n$ and $j=m$.
	\item Adjustment window condition: The warping path with minimum distance is searched on a restricted area on the alignment window to avoid significant timing difference between the two paths \cite{Sakoe1978}. The restricted area is given by $i-r \leq j \leq i+r$. 
	\item Slope constraint: This condition avoids significant movement in one direction \cite{berndt1994}. After $a$ steps in horizontal or vertical direction, it cannot move in the same direction without having $b$ steps in the diagonal direction \cite{Sakoe1978}. The effective intensity of the slope constraint can be defined as $P=b/a$. We chose $P=1$, which was reported as an optimum value in an experiment on speech recognition \cite{Sakoe1978}.  
	%\todo[]{how is an optimum value selected?}
\end{itemize}

In this paper, the distances between the time series are computed using the FastDTW package \cite{salvador2007}.

\subsection{Similarity matrices using DTW}
\label{sec:process}
%******************************************
Using DTW, we get a measure how different/similar any pair of time series $TS_{1}$ and $TS_{2}$ is. By comparing $N$ time series $TS_{1}, \ldots, TS_{N}$ with each other, we can generate similarity matrices whose entries are the distance between the two corresponding time series. Since DTW is not commutative, the resulting matrices are not symmetric. Consequently, the resulting similarity matrix for DTW requires $N^2$ computations instead of $N(N-1) / 2$. The similarity matrices can then be combined with a K-Nearest Neighbor classifier to inform us whether the time series corresponds to chatter or chatter-free cutting. During classification, data set is split into training set and test set. Indices of the training set and test set samples are found and then the distance matrix for training set and test set is generated by using the square distance matrix computed for all cases in the beginning.

However, before computing the similarity matrices, there are two necessary conditioning steps in addition to the ones described in Section~\ref{sec:experimental_setup}: 1) normalizing the time series to zero mean and standard deviation of one. This normalization is necessary to eliminate the effect of features with higher values on cost functions of classification algorithms \cite{Theodoridis2009}. And 2) subdividing the time series while maintaining the corresponding tagging. For the similarity matrix computations, the time series are divided into equal parts whose lengths are nearly $10000$ for the DTW computations so that we can decrease the computation time for the similarity measures. 
%For the DFD, the data set is instead divided into parts whose length is $1000$ since DFD is only defined for a pair of time series with equal lengths. 
%Before splitting the data set for the DFD implementation, the lengths of signals are rounded to the nearest integer multiple of $1000$ since dividing their original lengths by $1000$ yields a non-zero remainder.
%The reason for choosing $1000$ as the length of the divided time series is to avoid omitting too much data from original signal. 
%Although it may be possible to choose a number smaller than $1000$, this will increase the total number of the resulting time series thus increasing the overall computation time.

\section{K-Nearest Neighbor (KNN)}
\label{sec:k-Nearest-Neighbor}
%************************************

\begin{figure}[h]
\centering
\includegraphics[width=0.85\textwidth,height=.95\textheight,keepaspectratio]{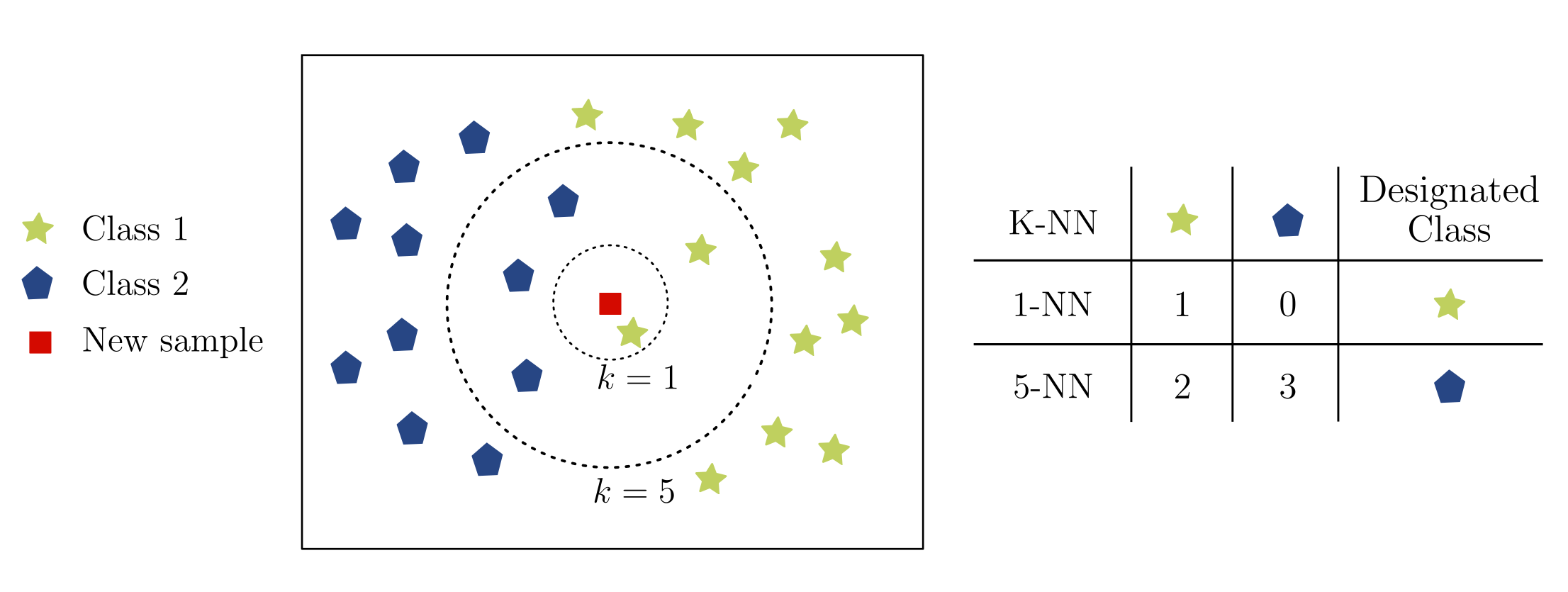}
\caption{K-Nearest Neighbor classification example for two-class classification.}
\label{fig:kNN}
\end{figure}

In this study, we used a K-Nearest Neighbor (KNN) algorithm to train a classifier. KNN is a supervised machine learning algorithm based on classifying objects with respect to labels of nearest neighbors \cite{dudani1976}. The `K' corresponds to the number of neighbors chosen to decide the label of newly introduced samples. Figure~\ref{fig:kNN} shows an example that illustrates the classification process with KNN. 

%---------------------------------
%---------------------------------

Specifically, Fig.~\ref{fig:kNN} assumes that we have two different classes for a classification problem denoted by pentagons and stars. Pentagons and stars belongs to the training set and the red square belongs to test set. Training set classification is performed by using the distance matrix between training set samples. The sample in training set is excluded and its label is designated with respect to the distances between remaining training set samples. On the other hand, test set classification is performed by using the distance matrix between training set and test set. When a new sample with unknown tagging is encountered (the square in the figure), we assign a tag based on the number of K nearest neighbors to each class. For instance, for the $1$-NN case, the closest neighbor is from the star class; therefore, the test sample is tagged as a star class. On the other hand, for the $5$-NN case, the test sample has two neighbors from the star class and three neighbors from from the pentagon class. Consequently, the label for the test sample is set as pentagon since there are more neighbors from this class than there are from the star class. In the case when the nearest neighbor number is selected even number and the there is a tie, the class label is assigned randomly with equal probability \cite{Horvath2001}. 
%\todo[inline]{Is there a standard way to break ties (as in when the number of neighbors from each class is equal for even values of K)?}

% The distance matrix for the training set is first computed, and it contains  the pairwise distances between its elements. 
% The distance matrix used for classification is then computed using the pairwise distances between the training set and test set. 
% Then, the K nearest neighbors for each of these samples in the training set are chosen by comparing the distances in training set distance matrix. 
% The class with the larger sample number in the neighbors is assigned to label of training sample. 
% The procedure is the same when we classify test set samples. 

\section{Results}
\label{sec:results}
%************************************
This section presents the results for the classification accuracy using our approach as well as current state-of-the-art methods in the literature.  
Specifically, Section~\ref{sec:dtw_results} compares the classification accuracy using the same data set for the similarity-based method to the WPT/EEMD methods \cite{Yesilli2019}, and the TDA-based results \cite{Yesilli2019a}. 
Further, Section~\ref{sec:transfer_learning_results} shows the transfer learning performance of the similarity measure method in comparison to the WPT/EEMD results \cite{Yesilli2019}. 

%**********************************************************
\subsection{Classification Results for Dynamic Time Warping(DTW)}
\label{sec:dtw_results}
%**********************************************************

Table~\ref{tab:results_comparison} compares the best classification scores obtained from WPT, EEMD, and the TDA-based methods to the results from DTW. 
The cells highlighted in green are the ones with the highest overall classification score, while those highlighted in blue represent results with error bands that overlap with the best overall accuracy in the same row.
A full list of the average classification scores and the corresponding standard deviations can be found in Tables~\ref{tab:DTW} for the DTW similarity matrices. 

\begin{table}[H]
\centering
\begin{threeparttable}
\caption{Comparison of results for similarity-based methods with their counterparts available in literature.}
\label{tab:results_comparison}
\resizebox{\textwidth}{!}{\begin{minipage}{\textwidth}
\begin{tabular}{|c|c|c|c|c|c|c|c|c|c|}
\hline
\multicolumn{1}{|c|}{} & \makecell{Similarity\\Measure} & \multicolumn{6}{c|}{Topological Data Analysis (TDA)} & \multicolumn{2}{c|}{Signal Processing} \\
\hline
\makecell{Stickout\\Length\\cm\\(inch)} & DTW &\makecell{Persistence\\Landscapes} & \makecell{Persistence\\Images} & \makecell{Template\\Functions} & \makecell{Carlsson\\Coordinates} & \makecell{Kernel\\Method} & \makecell{Persistence\\Paths} & WPT & EEMD\\
\hline
\makecell{5.08\\(2)}		& \cellcolor[RGB]{75,228,141}$\SI{99.5}{\percent}$  &$\SI{92.3}{\percent}$ & $\SI{76.1}{\percent}$&$\SI{89.6}{\percent}$ & $\SI{88.7}{\percent}$ & $\SI{74.5}{\percent}$*& $\SI{83.0}{\percent}$ & $\SI{93.9}{\percent}$ & $\SI{84.2}{\percent}$  \\
\hline
\makecell{6.35\\(2.5)}	& $\SI{75.8}{\percent}$ &$\SI{70.2}{\percent}$ & $\SI{66.7}{\percent}$&$\SI{87.7}{\percent}$ & $\SI{86.3}{\percent}$ & $\SI{58.9}{\percent}$ & $\SI{84.2}{\percent}$ & \cellcolor[RGB]{75,228,141}$\SI{100.0}{\percent}$ & $\SI{78.6}{\percent}$ \\
\hline
\makecell{8.89\\(3.5)}	& \cellcolor[RGB]{75,228,141}$\SI{94.6}{\percent}$  &$\SI{84.8}{\percent}$ & $\SI{83.5}{\percent}$&$\SI{84.3}{\percent}$ & \cellcolor[RGB]{204,229,255}$\SI{93.0}{\percent}$ & $\SI{87.0}{\percent}$ & $\SI{85.9}{\percent}$ & $\SI{84.0}{\percent}$ & \cellcolor[RGB]{204,229,255}$\SI{90.7}{\percent}$  \\
\hline
\makecell{11.43\\(4.5)}	& \cellcolor[RGB]{75,228,141}$\SI{81.5}{\percent}$  &$\SI{66.3}{\percent}$ & $\SI{66.4}{\percent}$&\cellcolor[RGB]{204,229,255}$\SI{78.1}{\percent}$ & $\SI{65.9}{\percent}$ & $\SI{59.3}{\percent}$ & $\SI{70.0}{\percent}$ & \cellcolor[RGB]{204,229,255}$\SI{78.8}{\percent}$ & \cellcolor[RGB]{204,229,255}$\SI{79.1}{\percent}$  \\
\hline
\end{tabular}
\end{minipage}}
\begin{tablenotes}
\small
\item *This result belongs to only the first iteration for the $5.08$ cm (2 inch) stickout case. 
\end{tablenotes}
\end{threeparttable}
\end{table}

Table~\ref{tab:results_comparison} shows that features based on WPT and DTW give the highest accuracy for the different stickout cases. 
For the $5.08$, $8.89$, and $11.43$ cm ($2$, $3.5$, and $4.5$ inch) stickout cases, DTW has the highest classification scoring accuracies of $99.5\%$, $94.6\%$ and $81.5\%$, respectively. 
On the other hand, feature extraction with WPT and RFE is the most accurate for the $6.35$ cm ($2.5$ inch) stickout cases scoring $100\%$. 
While the results from other methods are not the highest for any of the considered cases, some of them still lie within the error bars for the $8.89$ and $11.43$ cm ($3.5$ and $4.5$ inch) stickout cases. 
Specifically, EEMD and Carlsson Coordinates are within one standard deviation of the best results for the 8.89 cm ($3.5$ inch) case, while for the 11.43 cm ($4.5$ inch) case, Template Functions (Ref.~\cite{Perea2019}), WPT and EEMD results are also in the error bar of the best accuracy. 

Note that the results provided in Table~\ref{tab:results_comparison} are for two-class classification: chatter (which also include intermediate chatter), and no chatter. 
However, there might be interest in identifying intermediate chatter as part of a prediction algorithm that interferes before it develops into full chatter. 
Alternatively, inducing or sustaining intermediate chatter might be desirable for surface texturing applications. 
Figures~\ref{fig:tagged_data}b--d show that the power spectrum for chatter and intermediate chatter is very similar making the featurization in the frequency domain extremely challenging. 
However, Fig.~\ref{fig:tagged_data}a shows a clear difference between the two chatter regimes in the time domain. 
Therefore, it is more advantageous to extract features in the time domain for a three-class classification (chatter, intermediate chatter, and no chatter). 

\begin{figure}[htbp]
\centering
\includegraphics[width=0.9\textwidth,height=.85\textheight,keepaspectratio]{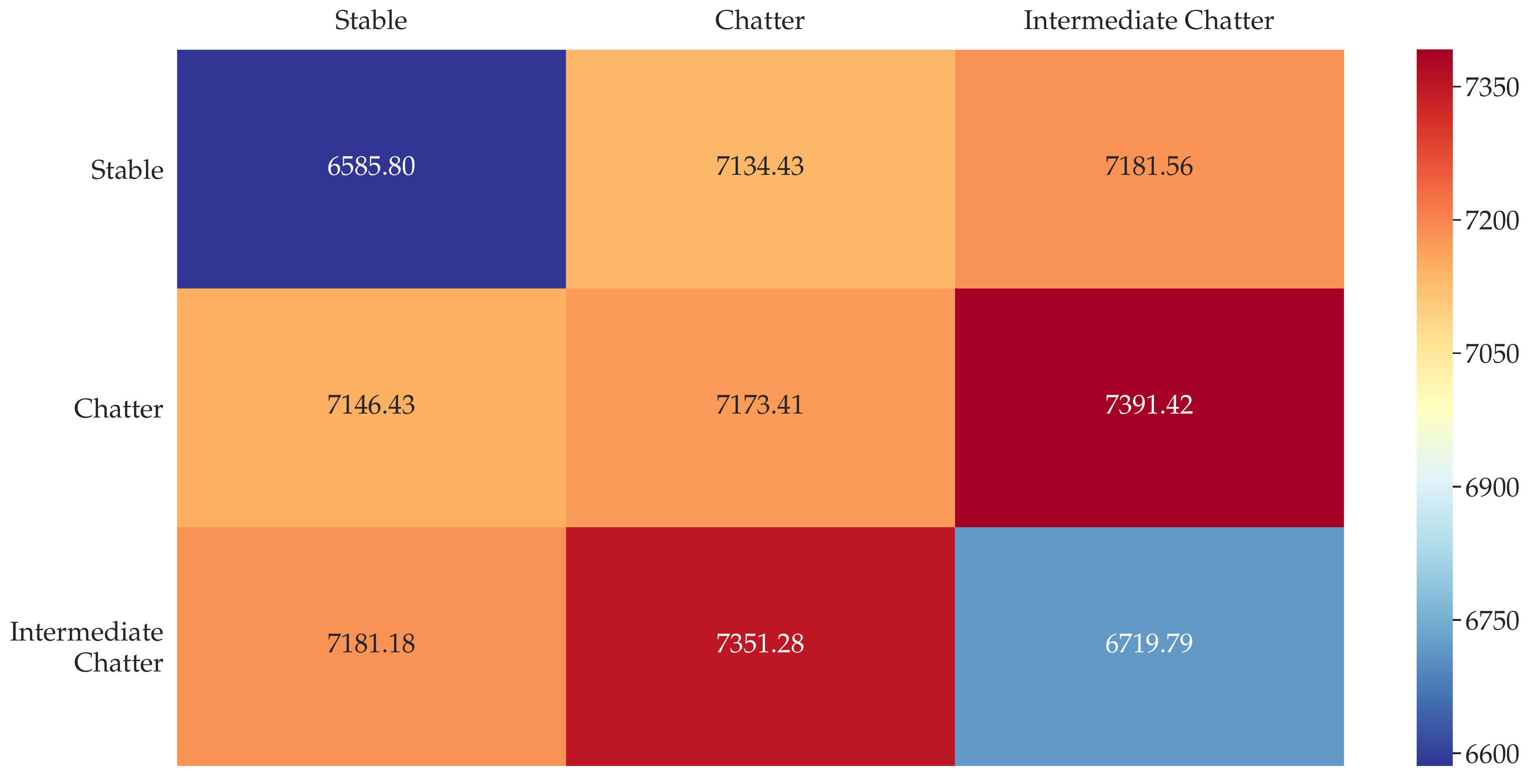}
\caption{The heat map that represents the average DTW pairwise distances for time series with different tagging. 
The number marked in each region is the average DTW distance of all the time series pairs in that region.
The used time series correspond to the case with 11.43 cm (4.5 inch) stickout length.}
\label{fig:4p5_inch_average_distance_matrix}
\end{figure}

Figure~\ref{fig:4p5_inch_average_distance_matrix} shows that DTW can differentiate between these chatter and intermediate chatter as evidenced by the high average distance between time series tagged as chatter and intermediate chatter. 
This figure corresponds to the cutting tests with stickout length of $11.43$ cm (4.5 inch), and each of the nine regions shows the average DTW distance between all the cases marked according to the row and the column labels in that region, e.g., the top right region reports the average DTW distance between the time series tagged as no-chatter versus those tagged as intermediate chatter. 
Therefore, looking at the regions that list the distances between intermediate chatter and chatter cases, we can see that the distances between chatter-chatter and chatter-intermediate chatter cases are quite different.
This confirms the ability of our approach to distinguish the differences between these two cases, provided a larger sample of intermediate chatter cases are available. 

%---------------
%---------------

As an example using our specific data set, we computed the distance matrices for each of the $6.35$ cm (2.5 inch) and the $8.89$ cm ($3.5$ inch) stick out cases, and applied the KNN classification algorithm to obtain the best $3$-class classification accuracy. 
Table~\ref{tab:three_class_classification_best} provides the best results of corresponding cases for three class classification (the full classification results can be found in Table~\ref{tab:three_class_classification_all}). 
Table \ref{tab:three_class_classification_best} shows that the DTW approach successfully distinguishes the three different classes. 

%---------------
\begin{table}[H]
\centering
\caption{Best results for three class classification with DTW approach, and the corresponding number of nearest neighbors used in the KNN algorithm.}
\label{tab:three_class_classification_best}
\begin{tabular}{|c|c|c|}
\hline
\makecell{Stickout Length \\ cm (inch)} & DTW & K-NN algorithm\\
\hline
\makecell{6.35 (2.5)}	&  $\SI{73.6 \pm 6.4}{\percent}$ & $1$-NN	  \\
\hline
\makecell{8.89 (3.5)}	&  $\SI{92.3 \pm 5.4}{\percent}$ & $4$-NN	 	\\
\hline
\end{tabular}
\end{table}
%---------------

%****************************************
\subsection{Transfer Learning Results}
\label{sec:transfer_learning_results}
\begin{figure}[htbp]
\centering
\includegraphics[width=0.9\textwidth,height=.85\textheight,keepaspectratio]{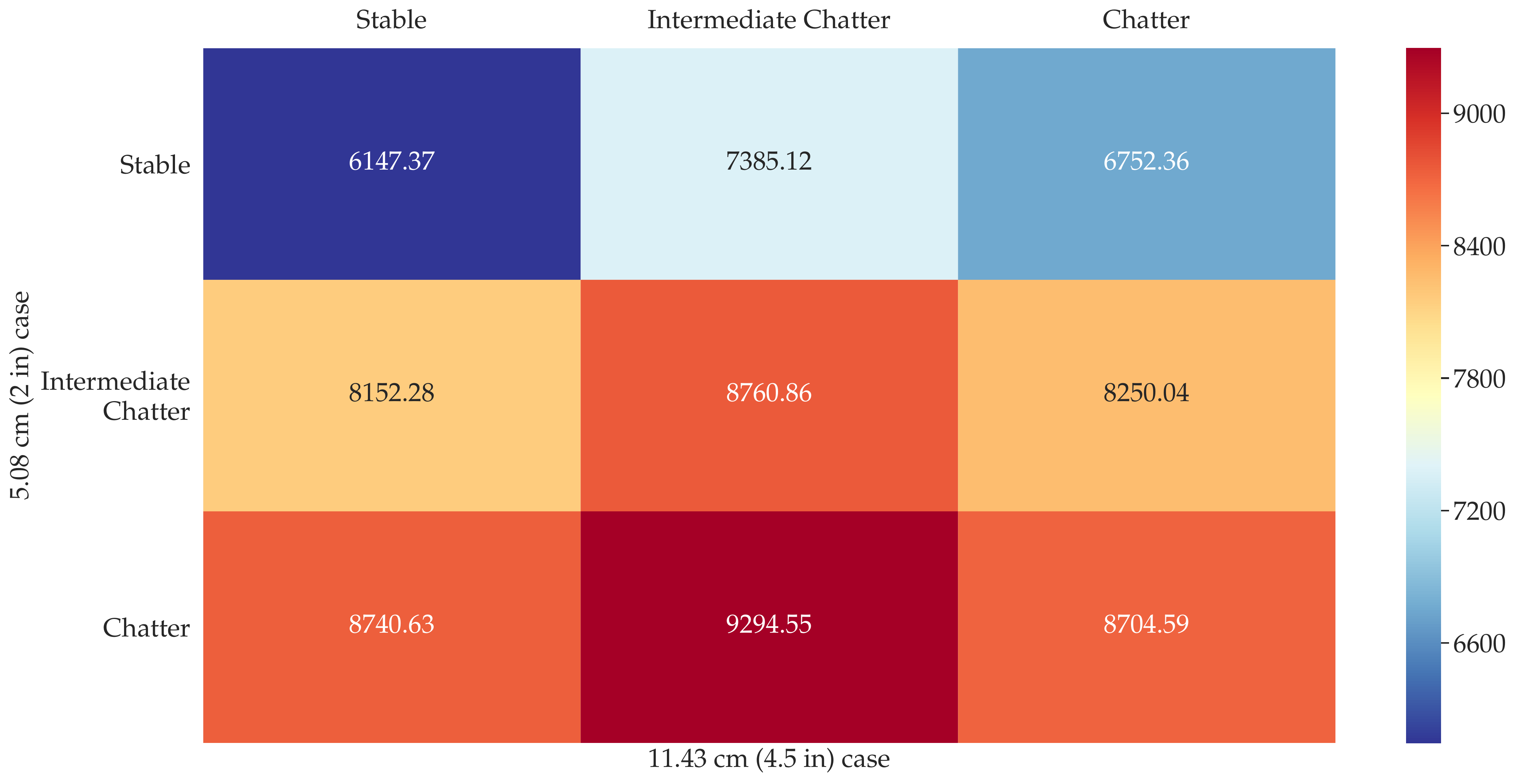}
\caption{The heat map that represents the average DTW pairwise distances between 5.08 and 11.43 cm (2 and 4.5 inch) cases. 
The number marked in each region is the average DTW distance of all the time series pairs in that region.}
\label{fig:2_4p5_inch_average_distance_matrix}
\end{figure}

%****************************************
An important question with practical implications is how well will a trained classifier perform on a real manufacturing center? 
This question is strongly related to the concept of transfer learning, i.e., the idea of training a classifier on some cutting configuration, and hoping that this classifier will be robust enough to the inevitable changes in the cutting systems dynamic parameters during manufacturing. 
In order to assess the capability of transfer learning using time series similarity measures, Fig.~\ref{fig:2_4p5_inch_average_distance_matrix} shows the average distance for each quadrant of the distance matrix between the $5.08$ and $11.43$ cm (2 and 4.5 inch) stickout lengths. 
It can be easily seen that cases with different labels, albeit from different cutting configurations, can still be differentiated since their pairwise distances are distinct. 
In contrast, Fig.~\ref{fig:tagged_data} shows that, for instance, chatter and intermediate chatter show similar frequency bands, which complicates extracting distinguishing features between chatter and intermediate chatter using frequency-based features.  

%---------------
%---------------

Motivated by Fig.~\ref{fig:2_4p5_inch_average_distance_matrix}, we performed transfer learning analysis using the DTW-based approach.  
This analysis involved training a classifier using the $5.08$ cm ($2$ inch) data, and testing it on the $11.43$ cm ($4.5$ inch) signals. 
The same analysis was repeated using the $11.43$ cm ($4.5$ inch) data as the training test, and the $5.08$ cm ($2$ inch) data as the test set. 
These two cases where chosen in order to test the effectiveness of the classifier in detecting chatter when the eigenfrequencies of the system change between two extremes.  
$67\%$ of the training set was used to train the classifier, and testing was performed using $67\%$ of the testing set. 
% The reason for using $67\%$ of both data sets is to obtain mean accuracy and deviation for classification. 
In each case, a KNN classifier was trained for $K \in \{1, 2, \ldots, 5\}$, and the highest resulting accuracy was listed in Table~\ref{tab:transfer_learning_comparison} (the full classification accuracies can be found in Table~\ref{tab:transfer_learning_DTW}).  
This table also compares the transfer learning results from DTW to its WPT and EEMD counterparts obtained from Ref.~\cite{Yesilli2019}.

%---------------
\begin{table}[H]
\centering
\caption{Transfer learning results comparsion between DTW, WPT and EEMD.}
\label{tab:transfer_learning_comparison}
\begin{tabular}{|l|c|c|c|c|}
\hline
\multicolumn{1}{|c|}{} & \multicolumn{2}{c|}{\shortstack{Training Set: 5.08 cm (2 inch) \\ Test Set: 11.43 cm (4.5 inch)}} & \multicolumn{2}{c|}{\shortstack{Training Set: 11.43 cm (4.5 inch) \\ Test Set: 5.08 cm (2 inch)}} \\
\hline
\multicolumn{1}{|c|}{Method} & Test Set & Training Set & Test Set& Training Set \\
\hline
DTW					&\cellcolor[RGB]{75,228,141}$\SI{78.2 \pm 2.4}{\percent}$ &$\SI{100.0 \pm 0.0}{\percent}$  &\cellcolor[RGB]{75,228,141}$\SI{87.7 \pm 4.2}{\percent}$ &$\SI{97.7 \pm 1.8}{\%}$ \\
WPT Level 1 &$\SI{59.4 \pm 14.6}{\percent}$ &$\SI{87.7 \pm 4.5}{\percent}$  &$\SI{62.9 \pm 21.5}{\percent}$ &$\SI{81.4 \pm 12.0}{\%}$ \\
EEMD        &$\SI{64.0 \pm 0.7}{\percent}$  &$\SI{93.8 \pm 0.3}{\percent}$  &$\SI{83.6 \pm 0.4}{\percent}$  &$\SI{77.0 \pm 0.6}{\%}$ \\
\hline
\end{tabular}
\end{table}
%---------------

Table \ref{tab:transfer_learning_comparison} shows that DTW outperforms WPT and EEMD in transfer learning. 
EEMD places second in transfer learning although it still performs poorly when the training set is the $5.08$ cm ($2$ inch) stickout case, while WPT gives the poorest results in the two considered examples. 
These results show that a classifier trained using the DTW approach retains good classification accuracies even if the dynamic parameters of the machining process deviate from their original values.  

Table~\ref{tab:time_comparison} compares the runtime for each different methods for chatter detection.
These comparisons were performed using a Dell Optilex 7050 desktop with Intel Core i7-7700 CPU and 16.0 GB RAM.
It can be seen that feature extraction with WPT and RFE is the fastest across all of the stickout cases, while EEMD is the second fastest. 
However, although DTW outperforms or matches the highest accuracies in three out of four stickout cases, it has the longest runtime. 
We point out that the WPT package that we used is optimized whereas the DTW package is not.

Nevertheless, many researchers have published on DTW optimization especially for data mining \cite{rakthanmanon2012}. 
These studies optimize the DTW algorithm to speed up runtime for sub-sequence search and distance matrix computation between time series and its query. 
The resulting algorithm allows performing fast queries on a single core machine in a very short time; thus, allowing using small consumer electronics to handle the data and possibly extract features from it in real-time. 
These kinds of optimization can enable in-process chatter detection on the cutting centers using the similarity measure approach using a classifier that is trained first offline and then loaded to a controller attached to the manufacturing center.

\begin{table}[H]
\centering
\caption{Time (seconds) comparison between similarity measure methods and its counterparts.}
\label{tab:time_comparison}
\resizebox{\textwidth}{!} {  % make the table a bit smaller so it doesn't spill out
\begin{tabular}{|c|c|c|c|c|c|c|c|c|c|}
\hline
\multicolumn{1}{|c|}{} & \makecell{Similarity\\Measure} & \multicolumn{6}{c|}{Topological Data Analysis (TDA)} & \multicolumn{2}{c|}{Signal Processing} \\
\hline
\makecell{Stickout\\Length\\cm\\(inch)} & DTW  &\makecell{Persistence\\Landscapes} & \makecell{Persistence\\Images} & \makecell{Template\\Functions} & \makecell{Carlsson\\Coordinates} & \makecell{Kernel\\Method} & \makecell{Persistence\\Paths} & WPT & EEMD\\
\hline
\makecell{5.08\\(2)}	  &833522  	& 98582 & 85601	&84364 & 84352	&1466747&118403	&\cellcolor[RGB]{75,228,141}116	&14540	\\
\hline
\makecell{6.35\\(2.5)}	&43044    & 118391& 23930	&23756 & 23752	&153759	&30563	&\cellcolor[RGB]{75,228,141}37	  &3372	  \\
\hline
\makecell{8.89\\(3.5)}	&13681    & 25527 & 11437	&11322 & 11320	&71553	&14401	&\cellcolor[RGB]{75,228,141}5	  &1583	  	\\
\hline
\makecell{11.43\\(4.5)}	&83420    & 80462 & 37966	&37623 & 37619	&542378	&48958	&\cellcolor[RGB]{75,228,141}7	  &3096  	\\
\hline
\end{tabular}
}
\end{table}

\section{Conclusion} 
\label{sec:conclusion}
This paper presents a novel method for chatter detection that combines similarity measures of time series via Dynamic Time Warping (DTW) with machine learning.
In this approach, the similarity of different time series is measured using their DTW distance, and any incoming data stream is then classified using the KNN algorithm. 
We test the classification accuracy of our approach using a set of turning experiments with four different tool stickout lengths, and we compare the resulting accuracy to two widely used methods: the Wavelet Packet Transform (WPT) and the Empirical Mode Decomposition, as well as to newly developed tools based on Topological Data Analysis (TDA).   
Our results in Table~\ref{tab:results_comparison} show that the DTW's classification accuracy matches or exceeds those of existing methods for three out of four different stickout cases.
This indicates that temporal features extracted using DTW are effective markers for detecting chatter in cutting processes. 
Topological Data Analysis (TDA) methods results are also close to the ones for similarity measures; however, one advantage for the DTW approach in comparison to TDA-based tools is that it does not require embedding the data into a point cloud, hence avoiding the complications associated with choosing appropriate embedding parameters. 

In addition to its high accuracy, the DTW-based approach operates directly on the time series thus bypassing the skill-intensive preprocessing step involved in the WPT and EEMD methods. 
Further, in contrast to deep learning techniques, such as neural networks, using DTW does not necessitate a large number of datasets for training. 
The feasibility of implementing DTW on a real cutting center is examined by investigating its transfer learning capabilities. 
Specifically, in contrast to WPT and EEMD, Table~\ref{tab:transfer_learning_comparison} and Fig.~\ref{fig:2_4p5_inch_average_distance_matrix} show that DTW possesses excellent transfer learning accuracy whereby the classifier is tested on a cutting configuration different from the one that was used for training. 
This is a significant advantages for our approach that still enables it to achieve high classification accuracies even if the system parameters (in our case the eigenfrequencies) shift during the process.

The DTW approach also successfully distinguishes between chatter and intermediate chatter as shown in Table~\ref{tab:three_class_classification_best}. 
These comparisons are difficult or impossible using only frequency domain features because the frequency content in these two cases is too similar while the time domain signals are different as evidenced by Fig.~\ref{fig:tagged_data} and the heat map shown in Fig.~\ref{fig:4p5_inch_average_distance_matrix}. 

Although Table~\ref{tab:time_comparison} shows that DTW clocks the longest runtime, we note that this slowdown is mostly related to the training phase because of the large number of necessary pairwise distance computations during the training/testing phase.
However, once the classifier is obtained, the necessary runtime for DTW will be significantly reduced because any new data is classified upon computing its pairwise distance with the training set, i.e., the only needed computation is equivalent to the evaluation of one row of the training/testing similarity matrix. 
We also note that Table~\ref{tab:time_comparison} does not include the time required for the manual preprocessing in the WPT and EEMD methods for choosing informative packets or decompositions. 
The actual time for these two methods is larger than the ones provided in the table depending on the number of the investigated time series and the skill of the person performing the preprocessing.
This is because WPT and EEMD require a cumbersome process for checking the frequency spectrum of the times series and examining the energy ratio of the wavelet packets of the time series. 
Furthermore, whereas the WPT algorithms are highly optimized, the Python scripts that we used in this paper for computing the DTW have little to no optimization. 
We hypothesize that further optimization using for example the ideas in \cite{rakthanmanon2012} will speed up the runtime for the similarity measures making them a viable option for on-machine chatter detection.

\section*{Acknowledgement}
This material is based upon work supported by the National Science Foundation under Grant Nos.~CMMI-1759823 and DMS-1759824 with PI FAK.
% \clearpage
% \newpage
\bibliography{Turning_DTW_ML}
% \clearpage
% \appendix
\appendixpage
\label{sec:appendix}

%---------------

\begin{table}[H]
\centering
\caption{Results obtained with DTW similarity measure method for 2, 2.5, 3.5 and 4.5 inch stickout case.}
\label{tab:DTW}
\begin{tabular}{|l|c|c|c|c|}
\hline
\multicolumn{1}{|c|}{} & \multicolumn{2}{c|}{5.08 cm (2 inch)} & \multicolumn{2}{c|}{6.35 cm (2.5 inch)} \\
\hline
\multicolumn{1}{|c|}{K-NN} & Test Set & Training Set & Test Set& Training Set \\
\hline
1-NN	&$\SI{99.4 \pm 0.4}{\percent}$&$\SI{99.4 \pm 0.4}{\percent}$&$\SI{74.7 \pm 6.6}{\percent}$&$\SI{76.8 \pm 4.5}{\percent}$\\
2-NN	&\cellcolor[rgb]{0.13,0.67,0.8}$\SI{99.5 \pm 0.5}{\percent}$&$\SI{100.0 \pm 0.0}{\%}$     &$\SI{70.7 \pm 5.2}{\percent}$&$\SI{78.6 \pm 3.5}{\percent}$\\
3-NN	&$\SI{98.7 \pm 0.7}{\percent}$&$\SI{99.6 \pm 0.3}{\percent}$&\cellcolor[rgb]{0.13,0.67,0.8}$\SI{75.8 \pm 6.5}{\percent}$&$\SI{85.6 \pm 2.1}{\percent}$\\
4-NN	&$\SI{98.9 \pm 0.8}{\percent}$&$\SI{99.7 \pm 0.3}{\percent}$&$\SI{69.7 \pm 8.4}{\percent}$&$\SI{78.6 \pm 2.8}{\percent}$\\
5-NN	&$\SI{97.9 \pm 0.9}{\percent}$&$\SI{98.9 \pm 0.7}{\percent}$&$\SI{73.7 \pm 7.4}{\percent}$&$\SI{80.9 \pm 3.1}{\percent}$\\
\hline
\multicolumn{1}{|c|}{} & \multicolumn{2}{c|}{8.89 cm (3.5 inch)} & \multicolumn{2}{c|}{11.43 cm (4.5 inch)}\\
\hline
1-NN	&\cellcolor[rgb]{0.13,0.67,0.8}$\SI{94.6 \pm 5.4}{\percent}$&$\SI{90.8 \pm 1.8}{\percent}$&\cellcolor[rgb]{0.13,0.67,0.8}$\SI{81.5 \pm 4.3}{\percent}$&$\SI{78.3 \pm 2.8}{\%}$\\
2-NN	&$\SI{93.0 \pm 5.0}{\percent}$&$\SI{100.0 \pm 0.0}{\%}$     &$\SI{78.2 \pm 4.6}{\percent}$&$\SI{96.5 \pm 2.1}{\percent}$\\
3-NN	&$\SI{90.6 \pm 5.7}{\percent}$&$\SI{91.4 \pm 2.6}{\percent}$&$\SI{74.1 \pm 5.2}{\percent}$&$\SI{83.7 \pm 3.6}{\percent}$\\
4-NN	&$\SI{91.1 \pm 5.1}{\percent}$&$\SI{92.2 \pm 2.4}{\percent}$&$\SI{77.5 \pm 5.4}{\percent}$&$\SI{86.7 \pm 2.4}{\percent}$\\
5-NN	&$\SI{91.3 \pm 5.3}{\percent}$&$\SI{90.5 \pm 2.3}{\percent}$&$\SI{76.7 \pm 6.0}{\percent}$&$\SI{79.7 \pm 2.9}{\percent}$\\
\hline
\end{tabular}
\end{table}
%---------------

%---------------
\begin{table}[H]
\centering
\caption{Transfer learning results for Dynamic Time Warping (DTW)}
\label{tab:transfer_learning_DTW}
\begin{tabular}{|l|c|c|c|c|}
\hline
\multicolumn{1}{|c|}{} & \multicolumn{2}{c|}{\shortstack{Training Set: 5.08 cm (2 inch) \\ Test Set: 11.43 cm (4.5 inch)}} & \multicolumn{2}{c|}{\shortstack{Training Set: 11.43 cm (4.5 inch) \\ Test Set: 5.08 cm (2 inch)}} \\
\hline
\multicolumn{1}{|c|}{K-NN} & Test Set & Training Set & Test Set& Training Set \\
\hline
1-NN & $\SI{76.8 \pm  2.5}{\percent}$   & $\SI{99.3    \pm 0.4}{\percent}$  & $\SI{84.7 \pm 3.7}{\percent}$   & $\SI{78.5 \pm  2.0}{\percent}$ \\
2-NN & \cellcolor[rgb]{0.13,0.67,0.8}$\SI{78.2 \pm  2.4}{\percent}$   & $\SI{100.0   \pm 0.0}{\percent}$  & \cellcolor[rgb]{0.13,0.67,0.8}$\SI{87.7 \pm 4.2}{\percent}$   & $\SI{97.7 \pm  1.8}{\percent}$ \\
3-NN & $\SI{77.0 \pm  2.6}{\percent}$   & $\SI{99.7    \pm 0.3}{\percent}$  & $\SI{81.0 \pm 2.3}{\percent}$   & $\SI{85.7 \pm  2.8}{\percent}$ \\
4-NN & $\SI{77.9 \pm  2.4}{\percent}$   & $\SI{99.8    \pm 0.2}{\percent}$  & $\SI{84.6 \pm 3.2}{\percent}$   & $\SI{87.1 \pm  2.7}{\percent}$ \\
5-NN & $\SI{76.5 \pm  3.1}{\percent}$   & $\SI{99.3    \pm 0.4}{\percent}$  & $\SI{79.7 \pm 3.1}{\percent}$   & $\SI{80.0 \pm  2.8}{\percent}$ \\
\hline
\end{tabular}
\end{table}
%---------------

%---------------
\begin{table}[H]
\centering
\caption{Three class classification applied with DTW approach.}
\label{tab:three_class_classification_all}
\begin{tabular}{|c|c|c|c|c|}
\hline
\makecell{} &\multicolumn{2}{c|}{2.5 inch} & \multicolumn{2}{c|}{3.5 inch} \\
\hline
\makecell{K-NN } & Test Set & Training Set & Test Set & Training Set \\ 
\hline
1-NN		& \cellcolor[rgb]{0.13,0.67,0.8}$\SI{73.6  \pm 6.4}{\percent}$    & $\SI{72.9 \pm  3.1}{\percent}$ & $\SI{91.6  \pm 5.0}{\percent}$    & $\SI{92.6 \pm  2.9}{\percent}$\\
2-NN		& $\SI{62.1  \pm 5.8}{\percent}$    & $\SI{73.9 \pm  2.8}{\percent}$ & $\SI{90.7  \pm 5.3}{\percent}$    & $\SI{92.8 \pm  2.4}{\percent}$\\
3-NN		& $\SI{65.3  \pm 4.8}{\percent}$    & $\SI{73.5 \pm  3.6}{\percent}$ & $\SI{89.8  \pm 9.0}{\percent}$    & $\SI{92.3 \pm  2.1}{\percent}$\\
4-NN		& $\SI{67.4  \pm 5.0}{\percent}$    & $\SI{73.5 \pm  2.5}{\percent}$ & \cellcolor[rgb]{0.13,0.67,0.8}$\SI{92.3  \pm 5.4}{\percent}$    & $\SI{91.9 \pm  3.0}{\percent}$\\
5-NN		& $\SI{68.5  \pm 5.7}{\percent}$    & $\SI{72.7 \pm  3.0}{\percent}$ & $\SI{90.7  \pm 5.8}{\percent}$    & $\SI{90.7 \pm  2.8}{\percent}$\\
\hline
\end{tabular}
\end{table}
%--------------

\end{document}